\def\etal{{\it et al.\/}}

\def\ie{{\it i.e.\/}}

\def\ltsima{$\; \buildrel < \over \sim \;$}
\def\simlt{\lower.5ex\hbox{\ltsima}}
\def\gtsima{$\; \buildrel > \over \sim \;$}
\def\simgt{\lower.5ex\hbox{\gtsima}}

\documentstyle[aaspp]{article}
\tighten
\singlespace

\begin{document}

\title{The survival of interstellar clouds against \\
Kelvin--Helmholtz instabilities}
\author{Mario Vietri $^1$, Andrea Ferrara$^2$, and Francesco Miniati$^3$}
\affil{
$^1$Osservatorio Astronomico di Roma \\ 00040 Monte Porzio Catone (Roma),
Italy \\ E--mail: vietri@coma.mporzio.astro.it \\
$^2$Osservatorio Astrofisico di Arcetri \\ 50125 Firenze, Italy 
\\ E--mail: ferrara@arcetri.astro.it \\
$^3$Dipartimento di Astronomia, Universit\`a di Firenze, \\
50125 Firenze, Italy 
\\ E--mail: miniati@arcetri.astro.it}

\begin{abstract}
{We consider the stability of clouds surrounded by a hotter confining 
medium with respect to which they are in motion, against Kelvin--Helmholtz
instabilities (KHI). In the presence of cooling, sound waves
are damped by dissipation. Whenever cooling times are shorter than 
sound crossing times, as they are in the normal interstellar medium, 
this implies that the instability generated
at the interface of the two media cannot propagate far from the
interface itself. To study how this influences the overall stability, 
first we derive an analytic dispersion relation for cooling media, 
separated by a shear layer. The inclusion of dissipation does not heal 
the instability, but it is shown that only a small volume around
the interface is affected, the perturbation decaying exponentially
with distance from the surface; this is confirmed by numerical simulations. 
Numerical simulations of spherical 
clouds moving in a surroundiing intercloud medium by which are pressure confined
show that these clouds 
develop a core/halo structure, with a turbulent halo, and a core in 
laminar flow nearly unscathed by the KHI. The related and previously reported 
``champagne effect'', whereby clouds seem to explode from their top sides, is cured 
by the inclusion of radiative losses.}

\end{abstract}

\section{Introduction}

Shearing flows are ubiquitous in astrophysics. A partial list includes 
Herbig--Haro objects (Stone, Xu \& Mandy 1995), 
disk plus hot corona accretion flows around compact
objects (Liang \& Price 1977), wind outflows from galaxies (Wang 1995), 
jets of all sorts and scales, bipolar flows, and even winds from the 
progenitor of
SN 1987a (Mc Cray \& Lin 1994). Besides these, there is the immense class
of two--phase media, like the ISM (Spitzer 1978), High Velocity Clouds
(Ferrara \& Field 1994, Wolfire \etal 1995), Broad Line Regions
of AGNs, protogalaxies (Silk \& Norman 1981), Ly$\alpha$ clouds
(Sargent \etal 1980, Giallongo 1995), which are required to give origin to just 
about everything astrophysical, from stars (Shu, Adams, \& Lizano 1987), through
Globular Clusters (Vietri \& Pesce 1995), to whole galaxies (Ikeuchi \& Norman
1991). 

Two--phase equilibria are common because they 
arise through a universal mechanism, thermal instability of radiative media 
(Field 1965, Balbus 1986), but, equally universally, they are subject 
to the shearing instability discovered by Kelvin
and Helmholtz (KHI from now on), whose potential danger 
to ISM clouds was noticed in computer simulations at least
twenty years ago (Woodward 1976) and is discussed in textbooks (Spitzer 1978). 

Possible stabilization of the KHI by magnetic effects was discussed at least
as early as 1955 (Michael 1955, Chandrasekhar 1961) and more recently studied in
detail by Miura (1984) and, particularly, by Malagoli, Bodo and Rosner
(1996). There can be little doubt that this stabilization mechanism is relevant 
to a large class of phenomena, including jets from AGNs and Galactic sources, 
and Giant Molecular Clouds. Yet, the magnetic fields in several important 
astrophysical situations, like Ly$\alpha$ clouds or protogalaxies, may not 
yet have had the time to grow to values necessary for stabilization, while in
other astrophysical situations the role played by the magnetic field is not
well--established. It thus seems worth its while to consider idealized situations
where magnetic fields are altogether neglected.

The severity of the KHI has been put into sharp focus by 
numerical simulations of shearing flows around unmagnetized clouds
(Murray, White, Blondin and Lin 1993, MWBL from now on)
which show that, in the absence of gravity, such clouds are disrupted
on a time scale comparable to the flow crossing time of the cloud. 
These authors considered a dense cloud moving inside
a potential well which also contains a tenuous, warm phase, such that the
two are in pressure equilibrium. The external gravitational field makes that 
the cloud speed $V$ be comparable to the warm phase sound speed $c_s$, so that
the relative motion is at the boundary between subsonic and supersonic,
$M\approx 1$. Then the cloud is subject to the KHI in a regime 
close to the incompressible one, where the growth rate is known and large.
Especially destructive, in these simulations, was the development of the
KHI on the largest scales. The physical explanation for this quick 
destruction (Doroshkevich and Zel'dovich 1981, DZ, from now on, Nittman, 
Falle \& Gaskell 1982, and Nulsen 1986) is the following: 
when the cloud is initially placed in a wind, a
stagnation point forms ahead of the cloud, whose high pressure accelerates 
the wind around the cloud; by Bernoulli's theorem, the pressure of the
wind is least where its speed is highest, \ie, at the top of the cloud.
This pressure minimum is overcome by the inner pressure of the cloud, 
which is being compressed in the direction of motion by the combined effect 
of the stagnation point and of its own inertia, and causes an overspilling
of the cloud from its top. We shall refer to this as the {\it champagne effect}. 

MWBL an DZ considered the possibility that these clouds were stabilized 
by self-gravity, showing that this requires a minimum mass $M_{min}$ 
which, compared to the cloud's Jeans mass $M_J$ is 
\begin{equation}
\label{zeldovich}
\frac{M_J}{M_{min}} = 0.15 \left(\frac{c_s}{V}\right)^3 \approx 0.1 
\end{equation}
(MWBL), 
independent of the density contrast $D \equiv \rho_{warm}/\rho_{cloud}$. 
Hence a cloud can only choose its own death: by self--gravitational collapse,
if $M \ga M_J$, or by disruption by KHI if $M \la M_J$. We shall refer to this
result as the {\it Zel'dovich paradox}. 

Is there an alternative to curing the champagne effect, without incurring the
Zel'dovich paradox? MWBL and DZ considered an adiabatic fluid; let us
drop this assumption, and consider a fluid subject to radiative losses.
It is well known that pressure waves in such a fluid are damped 
more often than amplified as they trudge along (Field 1965, Balbus 1986).
This suggests the following remedy: that clouds are (much) larger than 
the distance over which pressure waves are damped. In this way, when
depressurization is occurring at the top faces of the cloud, the inner
part will be slow or unable to respond to the matter outflow, and the
champagne effect may be slowed down, or possibly altogether removed. 

That cooling times are shorter than cloud crossing times may run
counter to intuition (Malagoli, Bodo and Rosner 1996), but this
certainly applies to the local ISM, where typical cooling times 
are $\approx 10^4\; yr$ and sound speeds $\approx 1\; km\; s^{-1}$
(Spitzer 1978), resulting in damping lengths of $\approx 3\times 10^{16}\;
cm$, much smaller than typical cloud radii. Thus pressure waves
carrying the Kelvin--Helmholtz instability inside the cloud are 
damped before crossing the whole cloud. 

We are thus trying to stabilize the KHI in a local sense. We do
not expect to find a dispersion relation for the Kelvin--Helmholtz 
modes which shows them to be stable. Rather, we expect to find a
situation where the KHI is always present, except that it is {\it unable to
propagate far from the interface}, and thus entrain in a catastrophic fate
the whole cloud. 

The flow of the argument and the plan of the paper are as follows: in
the next Section, we derive analytically the dispersion relation for
the shear layer between two idealized, radiative fluids. This problem is then tackled
numerically in Sec. 3, and it is shown that the KHI is not healed, but that it is
confined to a narrow region around the surface of discontinuity. 
In Section 4, we present numerical simulations of a realistic cloud embedded in a light 
wind with Mach number $\approx 1$, for both the adiabatic and the radiative 
cases. It is shown here that the inclusion of radiative effects cures the 
champagne effect. This section also contains the simulation of a cloud for a 
duration far exceeding the expected KHI timescale, in an effort to show that the
core/halo structure that radiative clouds develop to withstand the KHI
is statistically time--independent, and stable. A brief discussion in
Section 5, and a short summary in Section 6 close the paper.

\section{The dispersion relation}

We want to derive the dispersion relation for the Kelvin--Helmholtz instability
for two inviscid, but compressible and radiative fluids
separated by a tangential discontinuity, located at the $z=0$ plane
in the unperturbed problem. The zero-th order situation that we
envision thus has an upper fluid occupying the $z>0$ semispace, and 
moving rightward with speed $v = V > 0$, while
the lower one occupies the lower ($z<0$) semispace, and 
is moving leftward with speed $v = -V <0$. The two fluids are
both supposed to be in thermal and pressure equilibrium. Since the point we 
wish to make is rather general, we consider an idealized fluid whereby
radiative losses balance radiative gains (per unit volume) $\Lambda = 0$,
where the idealized forms of the net cooling function we choose is
\begin{equation}
\label{cool}
\Lambda = \rho^2 L(T) /m_H - \dot{E} \rho
\end{equation}
with $m_H$ is the mass per particle,
and we neglect chemical composition effects.
It is understood that both $L(T)$ and $\dot{E}$
can be different for the two fluids. The two fluids are in pressure equilibrium
with each other, which entails $\rho_+ c^2_{s,+} = \rho_- c^2_{s,-}$, where
$c_s$ is the sound speed, and we distinguish the upward, rightward moving 
fluid with a $+$ subscript, and the lower, leftward moving fluid with a $-$
subscript. Lastly, we shall also impose that each fluid is stable with
respect to the Field (1965) instability, which can be obtained by imposing
\begin{equation}
\label{q}
\frac{d \ln L(T)}{d \ln T} \equiv q > 1\;.
\end{equation}
The reason why we introduce a heating, albeit idealized, and the parameter
$q$ is to make sure that no local instability affects our analysis, and
that any further instability we may find is simply due to the instability
of the tangential discontinuity. 

Next we wish to perturb the equations of fluid dynamics pertaining to each fluid
separately; thus we supplement the usual equations
\begin{equation}
\label{continuity}
\frac{\partial \rho}{\partial t} + \nabla\cdot(\rho\vec{v}) = 0
\end{equation}
\begin{equation}
\label{impulse}
\frac{\partial \vec{v}}{\partial t} + \vec{v}\cdot\nabla\vec{v} = 
-\frac{\nabla p}{\rho}
\end{equation}
with the energy equation appropriate to a nonviscous, but radiative fluid
\begin{equation}
{3\over 2}\rho (\frac{\partial}{\partial t} + \vec{v}\cdot\nabla)  T-
T (\frac{\partial}{\partial t} + \vec{v}\cdot\nabla) \rho = \Lambda \;.
\end{equation}
We now consider perturbations of the form 
\begin{equation}
\label{perturb}
\delta\!X \propto e^{n t} e^{\imath k_x x} e^{\imath k_y y} f(z)
\end{equation}
where the still unspecified $z$--dependence derives from the fact that the
zero-th order problem is not homogeneous in the $z$--direction. 
We shall also use, for the convective derivative, the notation
\begin{equation}
\label{ncap}
N \equiv n + \imath \vec{k}\cdot\vec{V}
\end{equation}
for the same reason. Perturbing and linearizing the hydrodynamic equations 
in either fluid we find
\begin{equation}
\label{deltarho}
N \delta\!\rho + \rho\nabla\cdot\delta\!\vec{v} = 0
\end{equation}
\begin{equation}
\label{deltav}
N \delta\!\vec{v} = -\frac{\nabla \delta\!p}{\rho}
\end{equation}
\begin{equation}
\label{deltat}
\frac{3}{2} N \rho 
\delta\!T - T N \delta\!\rho = -(\Lambda_\rho \delta\!\rho
+\Lambda_T \delta\!T )
\end{equation}
where we have used
\begin{equation}
\Lambda_\rho \equiv \frac{\partial \Lambda}{\partial \rho}\;,\;
\Lambda_T \equiv \frac{\partial \Lambda}{\partial T}\;.
\end{equation}

We now determine the $z$--dependence of our problem. 
By straightforward computations we eliminate $\delta\!T$ between Eq.\ref{deltat}
and the equation of state $p = \rho T /m_H$, (Boltzmann's constant $k_B$ is
included in the definition of $T$, which is thus an energy)
to express $\delta\!p$ in terms
of $\delta\!\rho$, next we plug this $\delta\!p$ into Eq. \ref{deltav} and
then eliminate $\nabla\cdot\delta\!\vec{v}$ between Eq. \ref{deltav} and Eq.
\ref{deltarho} to obtain 
\begin{equation}
\frac{1}{f}\frac{d^2 f(z)}{d z^2} = k^2 + m_H N^2 \left[T + \rho
\frac{T N - \Lambda_\rho}{3\rho N/2  + \Lambda_T}\right]^{-1} \equiv k_a^2\;.
\end{equation}
The right hand side of this equation does not depend upon $z$, thus neither can 
the right hand one. We thus find for the $z$--dependence that
\begin{equation}
\label{fofz}
f = \left\{ \begin{array}{ll} e^{-k_a z} & \mbox{if $z>0$} \\
e^{k_a z} & \mbox{if $z<0$} \end{array} \right. 
\end{equation}
where we have implicitly assumed $Re(k_a) > 0$. 

It is convenient to rewrite $k_a$ somewhat. We have
\begin{equation}
\Lambda_\rho = \frac{\rho L}{m_H} \;,\; \Lambda_T = \frac{\rho^2}{m_H}
\frac{d L}{d T}\;,
\end{equation}
where we used the thermal equilibrium condition that applies to the zero-th 
order solution, {\it i.e.\/} $\Lambda_0 = 0$. Now we also define the net
cooling time
\begin{equation}
\label{tau}
\tau \equiv \frac{3}{2}\frac{m_H T}{\rho L(T)}
\end{equation}
and use the sound speed $c_s^2 = 5 T/(3 m_H)$, to obtain
\begin{equation}
\label{kadimensional}
k_a^2 \equiv k^2 + \frac{N^2}{c_s^2}\frac{N\tau + q}{N\tau + 3(q-1)/5}\;
\end{equation}

$k_a^2$ is complex--valued, since it depends upon $N$, implying that 
there will always be a $k_a$ such that $Re(k_a) > 0$, and this, in
turn implies (Eq. \ref{fofz}) that each perturbation penetrates only
a finite amount inside the cloud. This is true also 
for incompressible fluids, where it can be shown that the damping 
wavenumber $Re(k_a) \propto k$; in this case, clouds of finite size 
are always perturbed through their whole volume by KH perturbations
of suitably small wavenumbers. Dimensionally, this occurs because
there is no other wavenumber in this idealized problem; however, 
in the problem where cooling losses are included, there is another
wavenumber, $1/c\tau$. It will be shown later on, in Section 5, 
that this changes $Re(k_a)$ in a dramatic way: in fact, we shall
find that $Re(k_a) \ga 1/c\tau$ for {\it every} wavenumber, so that,
provided the cloud radius much exceeds $\approx c\tau$, most of
the cloud volume cannot be penetrated by the KH perturbations.

Incidentally, this also simplifies the problem at hand, by allowing us to 
neglect the usual need to determine which solutions for the growth rate 
belong to solutions with incoming or outgoing waves: once a solution
for $n$ has been identified, the corresponding spatial part is the only 
physical one, \ie, the root of Eq. \ref{kadimensional} such that the
perturbation decays away from the mid--plane. This argument also shows
that any solution of the dispersion relation, to be derived in the following,
is physical, \ie, it has a spatial part consistent with the idea that
only perturbations emanating from the surface of discontinuity
(as opposed to reaching it) ought to be included.

Now that we have determined the $z$--dependence of our problem, we turn
to the dispersion relation that we obtain by assuring that the pressure
is continuous across the perturbed interface. We consider a displacement
of the $z=0$ interface by an amount $\zeta$ in the $z$--direction, and we
assume for it a dependence
\begin{equation}
\zeta \propto e^{n t} e^{\imath k_x x} e^{\imath k_y y} 
\end{equation}
so that the $z$--velocity of a particle lying at the interface, $\delta\!v_z$, 
is given by 
\begin{equation}
\delta\!v_z = (\frac{\partial }{\partial t} + \vec{V}\cdot\frac{\partial }{
\partial \vec{x}} )\zeta = N \zeta\;.
\end{equation}
On the other hand, from Eq. \ref{deltav}, we find
\begin{equation}
\delta\!v_z = \pm \frac{k_a \delta\!p}{N \rho}
\end{equation}
where we have used Eq. \ref{fofz}, and the upper sign refers to the 
semispace $z>0$, the lower one to $z<0$. Eliminating $\delta\!v_z$ from the
two equations above, we find 
\begin{equation}
\label{reldisptrue}
\frac{N^2_+ \rho_+}{k_{a,+}} = - \frac{N^2_- \rho_-}{k_{a,-}}
\end{equation}
where, as stated above, the subscripts distinguish the fluids above
and below the tangential discontinuity. Since $v = V >0$ for $z>0$, and
$v=-V$ for $z<0$, we find 
\begin{equation}
N_+ = n+\imath\vec{k}\cdot\vec{V} \;,\; 
N_- = n-\imath\vec{k}\cdot\vec{V} \; .
\end{equation}
We must remember that $k_{a,\pm}$ depends upon $N_{\pm}$ through 
Eq. \ref{kadimensional}, and that the thermodynamic properties of the two fluids
($\rho, T, c_s, q$) may be different. We also define for convenience
\begin{equation}
\label{def2}
y \equiv \frac{n}{c_{+,s} k}\;,\; D\equiv \frac{\rho_+}{\rho_-}\;,\; 
r_\pm \equiv c_{+,s} k \tau_\pm \;,\; R \equiv\frac{c_{+,s}^2}{c_{-,s}^2}
\end{equation}

Eq. \ref{reldisptrue} is the dispersion relation we are seeking.
It depends upon cooling and heating processes only through the 
cooling time $\tau$; no simple analytic solutions are possible for 
the general case. 

First, we study the solutions of Eq. \ref{reldisptrue} in the
case of identical fluids, \ie,~$D=1$.  
The real
part of $y$ for the growing mode is shown in Fig. 1 for the particular
value $q=1.1$ (the solutions are not very sensitive to $q$ as long
as $q> 1$, as required for thermal stability). 
Deviations from the purely adiabatic case (corresponding to
$r \propto \tau \rightarrow \infty$) as $r$ is progressively decreased
are clearly seen. 
For small values of
$r$ the curves approach asymptotically the isothermal solution ($r=0$).
Stabilization of the KHI occurs at the critical Mach number
$M_c \ge \sqrt{2}$ for an adiabatic fluid, while in presence of radiative losses
there are no stable regions, even though the growth rate are noticeably smaller.
In the incompressible regime (small Mach numbers) it is $Re(y) \propto M$
independently of $r$, thus recovering the classical result;
however, deviations from linearity occur already at $M\sim 0.3$.
In summary,  high--Mach number flows are destabilized by
inclusion of radiation, and low--Mach number ones tend to be more stable.

Next, we study the influence of the density contrast $D$ on the stability of the
fluid. For the
{\it adiabatic} case, in Fig. 2 we show the real part of $y$ for the fastest growing 
mode, obtained by numerical solution of Eq. \ref{reldisptrue}, as a function of 
Mach number, and for several values of the density contrast $D$. The most 
apparent features are:
(i) the critical value $M_c$ decreases with increasing $D$, thus expanding 
the stability region well inside the subsonic domain;
(ii) the amplitude of the growth rate decreases with increasing $D$, and the 
location of its maximum shifts towards lower values of $M$. Note that for 
$D=0.002$, Mach numbers larger than 0.6 are absolutely stable.
The situation is considerably modified by the inclusion of radiative losses
as shown by Fig. 3, which refers to the parameters  $D=0.002$, $q_+=q_-=1.1$, 
$r_+=10^6$. The choice for $r_+$ is motivated by the fact that one can
prove that a background medium with the characteristics of the intercloud 
medium of the Galaxy can be considered, to the present purposes, as very close
to adiabatic ($r_+\gg 1$). As a general result, we find that cooling processes
exhacerbate the KHI: the peak of the growth rate located around $M\sim
0.5$ is amplified as the system departs from the adiabatic limit, and all 
Mach numbers become unstable, even though with moderate growth rates,
$Re(y) < 0.1$. 
Also, two different unstable modes are found, one of which becomes marginally 
stable ($Re(y)=0$) in the adiabatic case (compare with Fig. 2).

\section{Numerical experiments for the shearing layer}

To check and extend the above analytical results we have performed a large 
number of numerical simulations. To this aim we have used a 
hydrodynamic code based on a 2D TVD-MUSCL shock-capturing scheme (Van Leer 
1979, Jacobs 1991); the code is accurate to second order in both space and time.
The total grid in the highest resolution models presented below is 200 by 90 
zones. The code has positively passed all the standard hydrodynamical tests.

We have run two different types of experiments simulating a shearing layer
and a cloud engulfed by a wind; in the following we describe the shearing layer
experiment, leaving the discussion of the simulations of the cloud in a wind 
to the next Section.

In the first class of experiments the lower part of the grid is filled with a 
uniform fluid of density $\rho_-$ and sound speed $c_{-,s}$
moving leftwards at a Mach number  $M=V/c_{-,s}$ and the upper part is filled 
with a uniform fluid of density $\rho_+$ and sound speed $c_{+,s}$
moving rightwards at the same speed. The density contrast between the two 
fluids is $D=\rho_+/\rho_-=2\times 10^{-3}$ and they are supposed to be in 
pressure equilibrium ({\it i.e., } $D=1/R$, see Eq. \ref{def2}). The density 
at the interface is discontinuous and we assume the gas to evolve {\it
adiabatically}, {\it i.e., } $\Lambda=0$. We have imposed a sinusoidal 
initial perturbation of the velocity field, with wavelength 
equal to the grid size and 5\% amplitude; the perturbation is centered around
the interface on a strip encompassing 9 grid zones.

Fig. 4  shows a snapshot of the density and velocity field
for the two values of the shearing Mach number $M=0.5$ and $M=0.8$ for the
adiabatic case and $D =2\times 10^{-3}$,
at times $t=3.6\tau_d $, and $t=7.3\tau_d $, respectively; $\tau_d=\ell/V
D^{1/2}$ is the linear growth time of the instability.
For $M=0.5$ a structure similar to the  ``Kelvin's cat's eyes''  of the
KHI is clearly recognized, whereas for $M=0.8$ the initial perturbation is 
quickly damped leading to a stable state. Hence, the numerical simulations 
confirm the results obtained analytically: in the adiabatic case 
(and  $D=2\times 10^{-3}$) the KHI is stabilized for $M>0.6$. 

Fig. 5 shows a plot of the density and velocity field for the case of two
identical fluids, with $M = 0.5 $, in the adiabatic and radiative cases for the
two panels, respectively. The value of $r$ - the ratio between the cooling time
and the sound crossing time - for this simulation is $6\times 10^{-3}$. 
Here it can be seen that the inclusion of radiation
affects the shear layer stability exactly as predicted by the analytic
dispersion relation (Fig. 2): for this value of $M$ both  panels should
be unstable, but the lower one on a much longer time scale. In fact, the
results show that in the radiative case at $t > 20\tau_d$ there is no
trace of the instability yet, as expected.

\section{Numerical experiments for a cloud engulfed by a wind}

The same two--dimensional code mentioned in the previous section was used to 
study the more realistic problem of a cloud moving in a background medium by 
which it is pressure-confined. An important caveat is needed here: it
is well--known (Bayly \etal, 1988) that the existence of nonlinear
shear instability requires threee dimensions, so that a purely two--dimensional
simulation is necessarily limited in scope. It is nonetheless
useful as a first crack of the problem to check whether the ideas
discussed above do produce a stabilization. Further, three-dimensional
simulation shall be presented elsewhere.

Since it would be impractical to simulate
such a situation, due to the very large grids required to follow the cloud
motion, we study instead the related problem of a cloud engulfed by a wind
moving at a given Mach number $M=V/c_{+,s}$, where $c_{+,s}$ is the sound
speed in the wind. The subtleties related to this different
approach are discussed in detail by MWBL, who found that the physics of the
KHI is not changed by the possible onset of Rayleigh-Taylor instabilities,
which can be shown to grow on a much longer timescale.
Some care has to be devoted to the choice of the boundary conditions of the
problem. That at the bottom of the grid is {\it free-slip},
to exploit the symmetry of the problem (to save computational time we consider
only half a cloud); the conditions at the other three grid boundaries are
selected according to the speed of the flow.

The cloud is supposed to be a factor $500$ denser than the background
wind, and moving at $M=0.8$. We have selected this value of $M$ since from the
above discussed linear analysis it should correspond to a 
stable configuration (as it is indeed for the shearing layer even in the
nonlinear regime). In addition, we show the cloud evolution both 
in the adiabatic (Fig. 6) and in the radiative (Fig. 7) case.
The net cooling function adopted is the canonical one (see for details Ferrara
\& Field 1994) allowing for a two phase cloud-intercloud medium; thus both
the cloud and the wind are initially in thermal equilibrium.
The physical conditions for the cloud and the wind are $n_-=1$~cm$^{-3}$, 
$T_-=16$~K, $\ell\sim 40$~pc, and $n_+ = 2 \times 10^{-3} cm^{-3}$, $T_+ =
8000 K$, respectively; here $n_+$ is the wind number density, $T_+$ its 
the temperature and $\ell$ is the cloud radius. 
Using the previous parameters we can derive $r_+ \sim 30$ and $r_-\sim 0.4$.
As emphasized already, the wind is much less radiative than the cloud, and
this situation, as readily realized from a comparison of Figs. 2 and 3, leads
to a stronger KHI with respect to the case in which both fluids are adiabatic.
In the simulations, no gravity whatsoever is included.

From an inspection of Fig. 6 we see that the cloud becomes highly flattened
with sharp density and pressure gradients along the edges. A vortex sheet
develops in the rear part of the cloud initiating a mass loss from the cloud
itself.
A certain degree of arbitrariness is present when identifying the cloud at
each time step and therefore defining the mass loss.
For this reason we have plotted (thick line in Fig. 6) the isocontour 
encompassing  90\% of the initial mass. At the last calculated evolutionary
time ($t= 4 \tau_d$, where $\tau_d=\ell/V D^{1/2}$ is the linear growth time of 
the instability) the cloud appears to be extremely distorted. The effect which
we dubbed 'champagne effect' is patently obvious.

Fig. 7 presents the analogous simulation for the radiative case. The
evolution is similar, but the instability appears less pronounced, as
can be appreciated from the velocity structure of the vortex. The cloud
appears to be more rounded and the density gradients less steep. 

The most obvious feature of Fig. 7, when compared with Fig. 6, is that the
innermost density contours are much less strongly distorted; actually, the 
cloud's core does not seem to have expanded at all. This visual impression is 
borne out by Fig. 8, where we plotted the time evolution of the volumes
occupied by the innermost $25\%, 50\%, 75\%$ and $90\%$ of the mass,
for both the adiabatic and the radiative simulations. From these it
is apparent that the outermost mass shells expand in both simulations,
albeit by different amounts, while the innermost ones behave in a
radically different way: they expand for the adiabatic case, and they
contract for the radiative one. Hence, the visual impression of 
disruption that MWBL attributed to their simulations (here effectively
reproduced in our Fig. 6) is strongly reduced by inclusion of radiative
losses. 

The next obvious question is to see whether the stratification of the
radiative cloud seen in Fig. 7 is in some sense stable. In Fig. 9
we show the time evolution of a special simulation which we carried out
for a rather long time, $t = 7.47 \tau_d$. The total time span of this
simulation is limited neither by computational power, nor by our patience,
but by the fact that, in the adopted reference frame, the cloud falls
off the right hand edge because of the net deceleration that the wind 
imparts to the cloud center of mass. 
Extension of Fig. 9 to longer 
timescales requires a prohibitively large computational grid.
Still, Fig. 9 manages to convey the obvious feeling that the cloud's 
modification has by and large stopped at $t \sim 6 \tau_d$, and from then 
on the cloud is essentially in a statistical equilibrium, with a 
well--developed core/halo structure.

Fig. 9 contains our major result, that sheared clouds are
stabilized against the KHI by radiative effects, without any appeal
to self--gravity which is here completely neglected.

\section{\label{discussion}Discussion, with a solution of Zel'dovich paradox}

It is well--known that, when the detailed structure of the shear layer 
is considered, modes with wavelengths shorter than the width of the
region over which there is a significant velocity gradient are
stabilized against the KHI (Chandrasekhar 1961). Accordingly, Nulsen (1982)
has argued that all modes with wavelength smaller than a cloud radius are also
stabilized. However, MWBL's simulations, and Fig. 7 above bear witness
to the strength of the remaining instability. We thus concentrate on these,
longer wavelength modes, and consider Eq. \ref{kadimensional} on the cold,
dense side of the interface. Since $k c_{-,s}\tau_- \ll 1$ 

\begin{equation}
\label{help}
Re(k_a) \approx \frac{N}{c_{-,s}}\left(\frac{N\tau_- + q}{N\tau_- + 3(q-1)/5}
\right)^{1/2}\;.
\end{equation}
We need to treat separately the two cases $N\tau_- \gg 1$ and $N\tau_- \ll 1$.

\subsection{ Case 1: $N\tau_- \gg 1$}  

In this case, we find 
\begin{equation}
Re(k_a) \approx \frac{Re(n)}{c_{-,s}} + \frac{2q+3}{10}\frac{1}{c_{-,s}\tau_-}\;.
\end{equation}
Here we can take $Re(n) \la k V D^{1/2}$ ($D$ is the density contrast
between warm and cold phases, Eq. \ref{def2}), the value for incompressible
fluids, because we know from Fig. 3 that the real instability timescale is 
longer for the compressible, radiative case. Then it can be shown that
the second term dominates the first one if the cloud radius $\ell$ much exceeds
$\ell_{c1}$, where 
\begin{equation}
\ell_{c1} = \frac{10 D^{1/2}}{2q+3} 2\pi V\tau_-\;.
\end{equation}
From the condition of pressure balance between the two phases, and that of
barely sonic motion, it can be seen that the above condition is equivalent, 
in order of magnitude, to $k c_{-,s} \tau_- \ll 1$. We then obtain 
\begin{equation}
\label{ka1}
Re(k_a) \approx  \frac{2q+3}{10}\frac{1}{c_{-,s}\tau_-}\;.
\end{equation}
From the above we find that $Re(k_a) \gg 1$ because of our assumption
$k c_{-,s}\tau_- \ll 1$, so that perturbations are damped in this radius range.
Thus all clouds with radius in the range 
\begin{equation}
\label{range1}
2\pi c_{-,s}\tau_- < \ell < 2\pi V\tau_-
\end{equation}
are stabilized against KHI. 
 
\subsection{ Case 2: $N\tau_- \ll 1$}  

The square root in Eq.  \ref{help} can now be approximated to yield
\begin{equation}
\label{help2}
Re(k_a) \approx \frac{5 q}{3(q-1)} \frac{Re(n)}{c_{-,s}} + \frac{25(2q+3)}{9(q-1)^2}
\frac{k^2 V^2\tau_-}{2c_{-,s}}\;.
\end{equation}
Again using for $Re(n)$ the value for incompressible fluids, we see that the
second term dominates the first one for $\ell \ll \ell_{c2}$, where
\begin{equation}
\ell_{c2} = \frac{2q+3}{6q(q-1) D^{1/2}} 2\pi V\tau_-\;.
\end{equation}
In this case, Eq. \ref{help2} can be approximated by retaining only the second 
term, and it can be shown that $Re(k_a) \gg 1$. Again, we find that 
all modes with radius in the range
\begin{equation}
\label{range2}
2\pi V\tau_- < \ell < \frac{2q+3}{6q(q-1) D^{1/2}} 2\pi V\tau_-
\end{equation}
are stabilized against the KHI. Together, Eq. \ref{range1} and \ref{range2}
imply that all modes in the range
\begin{equation}
\label{range}
2\pi c_{-,s}\tau_- < \ell < \frac{2q+3}{6q(q-1) D^{1/2}} 2\pi V\tau_-
\end{equation}
are stabilized against the Kelvin--Helmholtz modes. It should be noticed
that the above estimate is pessimistic, because, in deriving it, we have used
the instability rate for incompressible fluids, rather than the slower one
for realistic, compressible and radiative fluids of Fig. 3.
For a numerical evaluation, we use the standard ISM parameters (Spitzer 1978):
$\tau_- \approx 10^4 \; yr$, $c_{-,s} \approx 1\; km\; s^{-1}$, $V \approx
10\; km\; s^{-1}$, to obtain that all clouds in the range 
$0.03 {\rm pc}< \ell < 33$~pc are stabilized.
In particular, our simulations use the numerical values $\tau_- = 7 \times 10^5 \;
yr$, $c_{-,s} = 0.4\; km\; s^{-1}$, $V = 8\; km\; s^{-1}$ which implies that
the cloud size we have adopted ($\ell=40$~pc) is well into the corresponding
range of stability $2 {\rm pc}< \ell < 3000$~pc. 

We can now discuss the solution of the Zel'dovich paradox. We compare $\ell_{c2}$
with the Jeans length for a cloud $\ell_J$:
\begin{equation}
\frac{\ell_{c2}}{\ell_J} \approx 1
\end{equation}
for the standard ISM parameters. This implies that all clouds with mass 
$10^{-6} M_J < M < M_J$ are stabilized. This ought to be contrasted with Eq. 1.

The above argument shows that clouds are stable, at least for a time
comparable to the dynamical time scale $\tau_d$. 
The larger question of the persistence of this flow, \ie, how long 
the cloud with the core/halo structure will persist, can be dealt with
here only approximately. The persistence clearly depends upon two factors:
first, the net rate (gains minus losses) of mass entrainment of the cloud by 
the hotter medium, which we cannot determine here but which is surely smaller 
than $\la \pi \ell^2 \rho_h V$, where $\rho_h$ is the
density of the warm phase. Second, the kinetic energy losses due to the 
acceleration (in the cloud system of reference) of the warm phase matter.
Both arguments lead to a deceleration timescale $\tau_d$ of order
\begin{equation}
\tau_d = \frac{4}{3} \frac{R}{V} \frac{\rho_c}{\rho_h} \approx 300 \frac{R}{V}
\;,
\end{equation}
which shows the survival time to be a few hundred times the dynamical timescale,
$\approx 3\times 10^8\; yr$. Most likely, this time is long enough for
other factors like collisions to become important. In summary, the inclusion 
of radiative effects has eliminated the Zel'dovich paradox, leaving a range 
of $\approx 6$ orders of magnitude of mass within which clouds are stable with 
respect to both self--gravity and KH modes. 

Another caveat that is worth discussing is that it is not at all obvious 
that clouds in the ISM, and the confining gas should be in thermal equilibrium,
but this only strengthens our arguments. It seems in fact that most clouds
are slowly cooling down, with unreplenished losses. When the equation of state
softens as the pressure waves trudge along, they are damped: they
put more work into compressing the ISM than is returned to them because of 
radiative losses. So, by considering a situation of thermal equilibrium, we have
put ourselves into the least favorable conditions. That some kind of stabilization
is however achieved under these circumstances seems to us a sufficiently general
point worth making. 

The last point we wish to make is that the above discussion closely parallels
that made in textbooks (Landau and Lifshitz 1987) for the development
of the phenomenon of separation in incompressible fluids, whereby turbulent,
rotational eddies cannot penetrate the laminar flow region, with a 
skin depth $\propto k^{-1}$. This is exactly similar to the discussion above,
except for the different dependence of the skin depth upon wavenumber.
Still, the parallel suggests where, ultimately, the shear energy will go:
in turbulence of a thin layer around the surface of separation, without 
disturbing the laminar flow of the remaining region, a conclusion which
we cannot, formally, extrapolate neither from our linear computations, 
nor from our coarse numerical simulations.

\section{Summary}

In this paper we have tried to show what follows:
\begin{itemize}
\item the KHI is not stabilized by inclusion of radiative losses;
\item however, the instability is contained within a small volume around the
surface of discontinuity;
\item thus stabilization occurs because cooling/heating timescales are 
shorter than dynamical ones: the instability does not extend much beyond the 
interface, and thus does not penetrate the main bodies of fluid involved;
\item this is borne out by analytical and numerical treatment of the shear 
layer;
\item clouds with radiative losses are stabilized against the 
`champagne effect';
\item this happens through the development of a core/halo structure 
whereby the halo is turbulent but incapable of exporting this turbulence
into the denser core, where laminar flow still prevails;
\item this defense mechanism, based only on thermodynamic and hydrodynamic
details, ought to be reasonably universal, and apply to the large range
of physical conditions covered by two--phase equilibria.
\end{itemize}
Our major result is visible in Fig. 9, where the evolution of a cloud 
for a time much longer than the timescale on which the KHI is expected to
operate has been followed, showing that the radiative cloud has reached a 
statistical equilibrium. 

\vskip 2truecm
Thanks are sue to an anonymous referee for comments that greatly improved the
manuscript.

\newpage

\begin{figure}

\caption{
Nondimensional growth rates $y=n/c_s k$ as a function of the
flow Mach number (unstable mode) for the identical fluids ($D=1$) case with
$q=1.1$. The numbers show the value of the parameter $r$ (the ratio between
cooling and sound crossing time of the perturbation) for the radiative cases;
the label $a$ denotes the adiabatic case.
}

\caption{
Nondimensional growth rates $y=n/c_{+,s} k$ as a function of the
flow Mach number (unstable mode) for the different, adiabatic  fluids case.
The numbers show the value of the parameter $D$, the density contrast between
the two fluids.
}
\caption{
Nondimensional growth rates $y=n/c_{+,s} k$ as a function of the flow Mach 
number (for the two unstable modes) for the different, radiative fluids 
case with a density contrast $D=0.002$, $q_+=q_-=1.1$, and $r_+=10^6$.
The numbers show the value of the parameter $r_-$ for the two unstable modes.
}

\caption{
Advanced stages of the evolution of the shearing layer with $D=0.002$. 
The two cases differ for the fluid Mach number: $M=0.5$, time $t= 3.6\tau _{d1}$
(upper panel) and $M=0.8$, $t= 7.3\tau _{d2}$ (lower panel). 
Due to the different Mach number,  
the corresponding dynamical times $\tau _{d1}$ and $\tau _{d2}$
(see Sec. 3) are different. The whole rectangular grid is presented here;
logarithmic density levels (contours) and velocity field (arrows)
are plotted. Density values range from $n _{min}=.8 $~cm$^{-3}$
to $n _{max}=577.5$~cm$^{-3}$;
velocity values range from $v _{min}=.012$ to 
$v _{max}=.88$ (in units of the isothermal sound speed in the rarefied gas).
}
\caption{
Advanced stages of the evolution of the shearing layer with $D=1$ and Mach 
number $M=0.5$; the adiabatic (upper panel) and radiative $r\sim 10^{-3}$ 
(lower panel) cases are shown at times $t= 6.76\tau _{d}$ and $t= 22.8\tau
_{d}$, respectively.
logarithmic density levels (contours) and velocity field (arrows)
are plotted. For the adiabatic case density values range from
$n _{min}=.77 $~cm$^{-3}$ to $n _{max}=1.05$~cm$^{-3}$; for the radiative case
density values range from $n _{min}=1.0 $~cm$^{-3}$ to $n _{max}=1.02$~cm$^{-3}$.
}
\caption{
Adiabatic evolution of a pressure confined cloud engulfed by a wind moving at 
a Mach number $M=0.8$.; the density contrast is $D=0.002$. 
Evolutionary times shown are (a) 0, (b) 1.32, (c) 2.43, 
(d) 3.53$\tau_d$, where $\tau _d$ is the KHI characteristic time scale defined 
in  Sec. 3. Contours correspond to logarithmic density levels and arrows 
show velocity field. The thick line in (d) represents the contour containing 
$90\%$ of the initial cloud mass. The minimum and
maximum densities (in units of the wind density) present are
(a) 1-500, (b) 0.46-673, (c) 0.49-965 and
(d) 0.4-514. The velocity range (in units of the isothermal sound speed 
in the wind) are (a) 0-1.03, (b) 0.001-1.86, (c) 0.005-1.86  and (d) 0.007-1.89. 
The grid is rectangular and cylindrical symmetry is assumed. 
}
%\end{figure}
%\newpage
%\begin{figure}
\caption{
Radiative evolution of a pressure confined cloud engulfed by a wind moving at 
a Mach number $M=0.8$. Evolutionary times shown are (a) 0, (b) 1.27, (c) 2.44, 
(d) 3.65$\tau_d$, where $\tau _d$ is the KHI characteristic time scale defined 
in  Sec. 3. Contours correspond to logarithmic density levels and arrows 
show velocity field. The thick line in (d) represents the contour containing 
$90\%$ of the initial cloud mass. The minimum and
maximum densities (in units of the wind density) present are
(a) 1-500, (b) 0.6-1139, (c) 0.53-1066 and
(d) 0.64-2053. The velocity range (in units of the isothermal sound speed 
in the wind) are (a) 0-1.03, (b) 0.001-1.55, (c) 0.0007-1.61 and (d) 0.002-1.53.
The grid is rectangular and cylindrical symmetry is assumed. 
}
\caption{Time evolution of the fractional volumes
occupied by the innermost $25\%, 50\%, 75\%$ and $90\%$ of the initial mass,
for both the adiabatic (open squares) and the radiative (open stars)
cases shown in Fig. 7
}
\caption{Same as Fig. 7 but for  later evolutionary times: (a) 4.47, (b) 5.28,
(c) 6.44, (d) 7.47$\tau_d$ 
}
\caption{Isobaric contours corresponding to the same case as shown in Fig. 7
}
\end{figure}
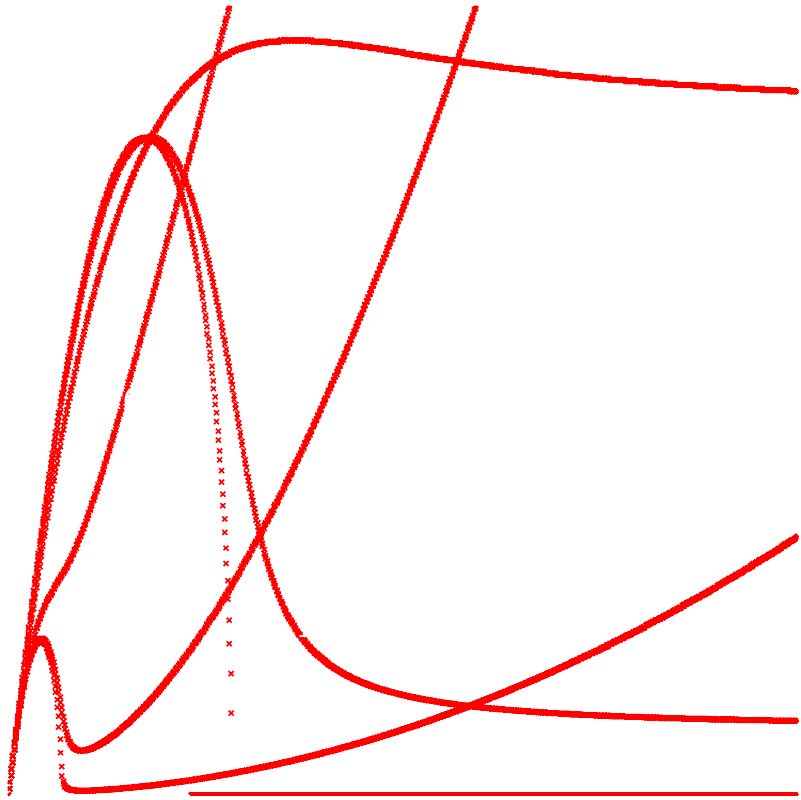
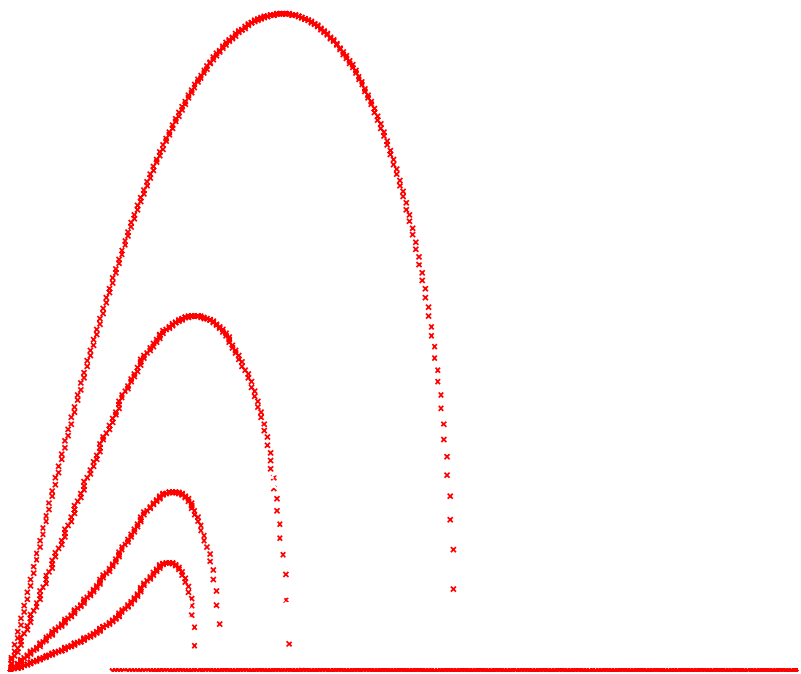
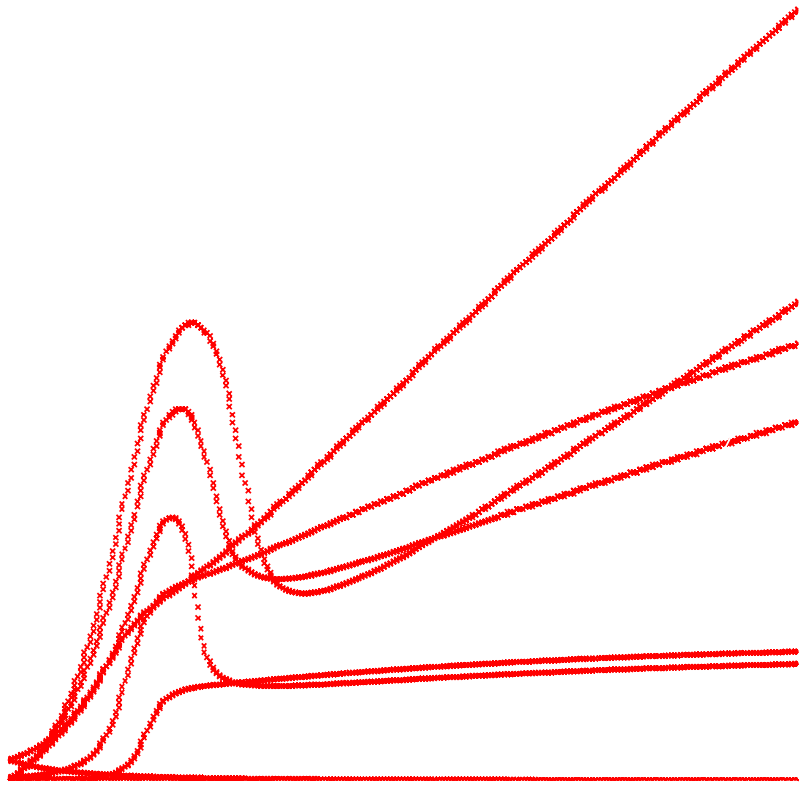
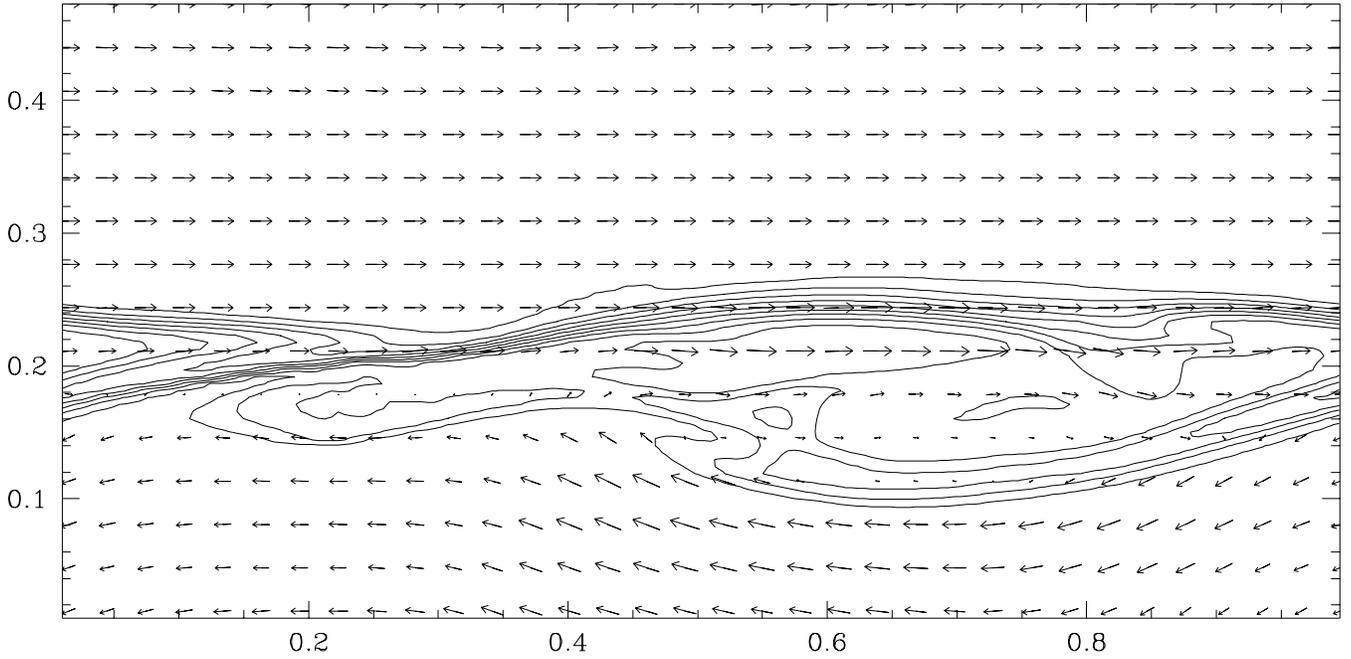
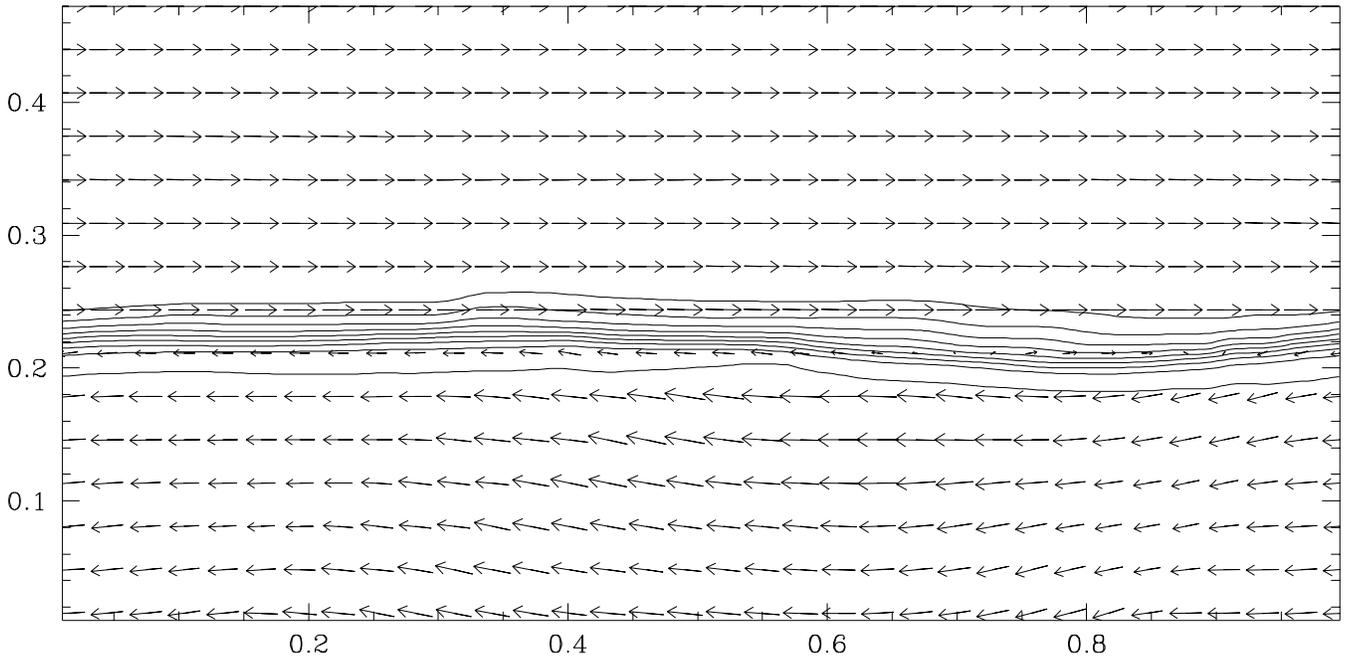
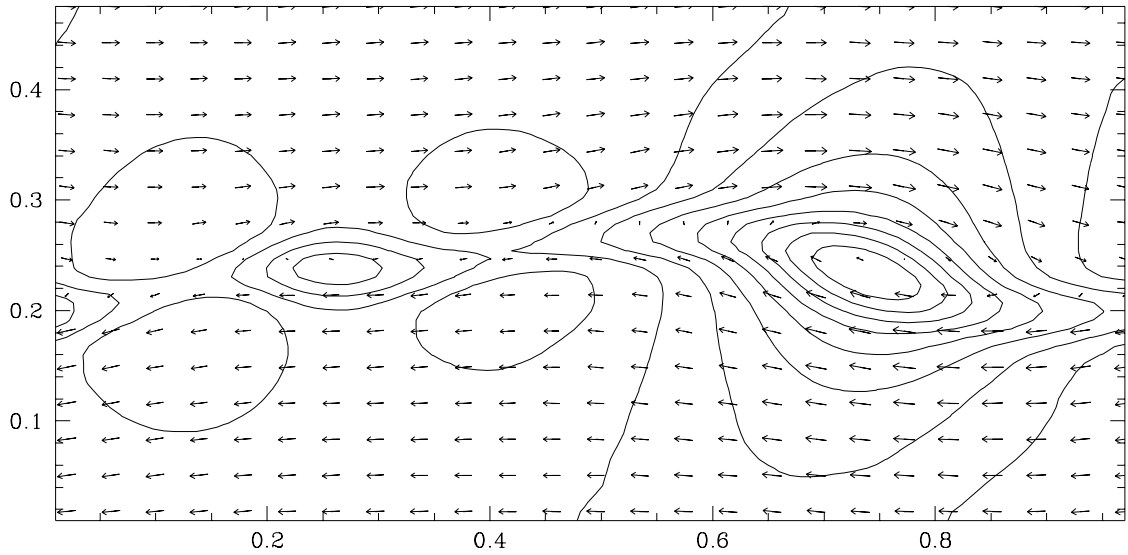
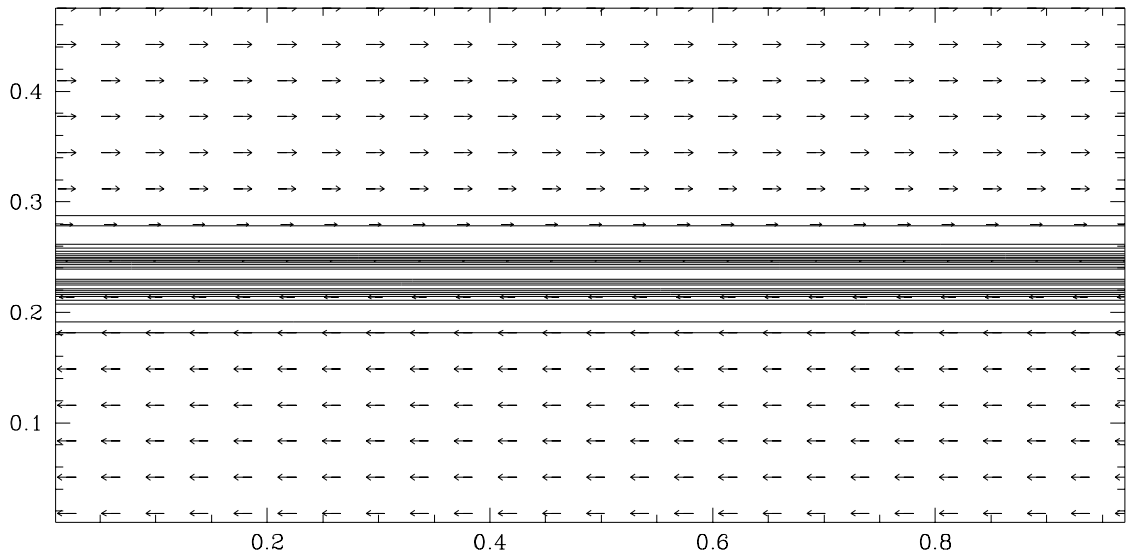
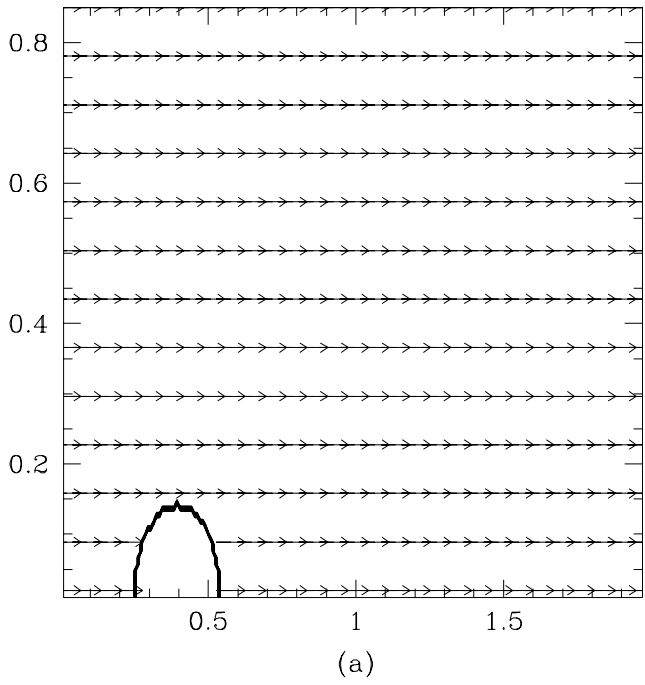
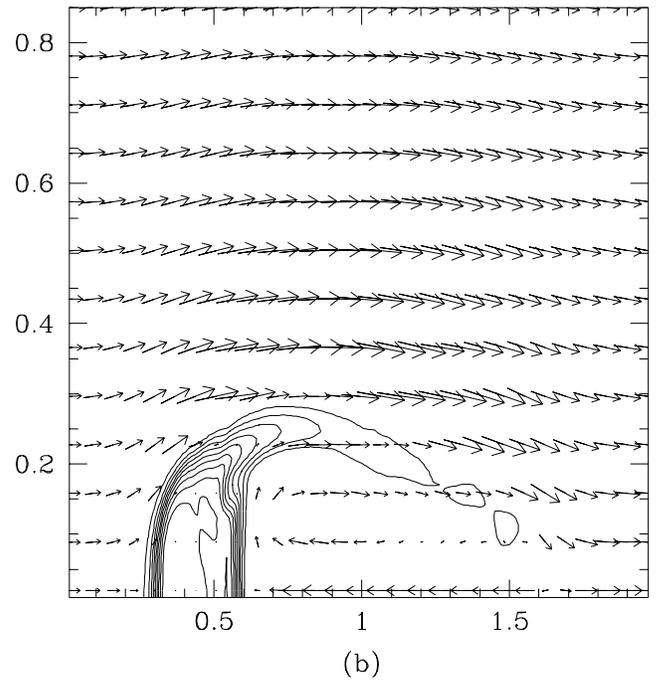
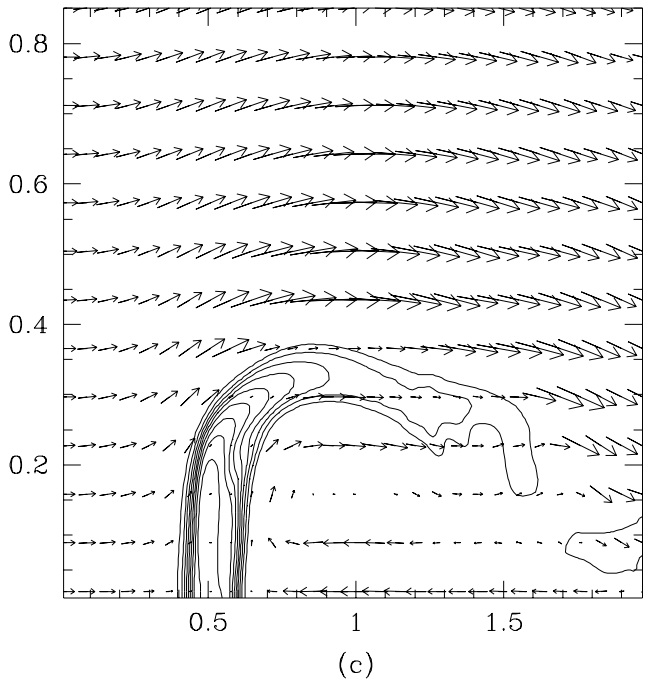
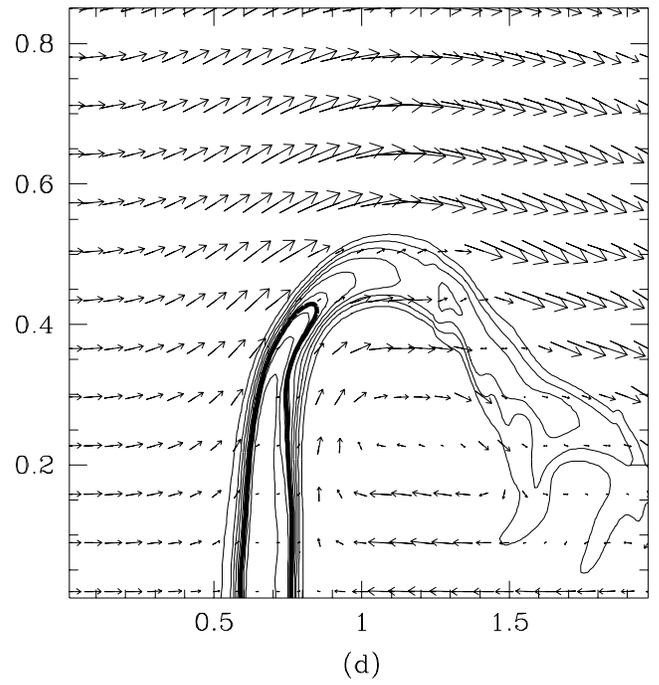
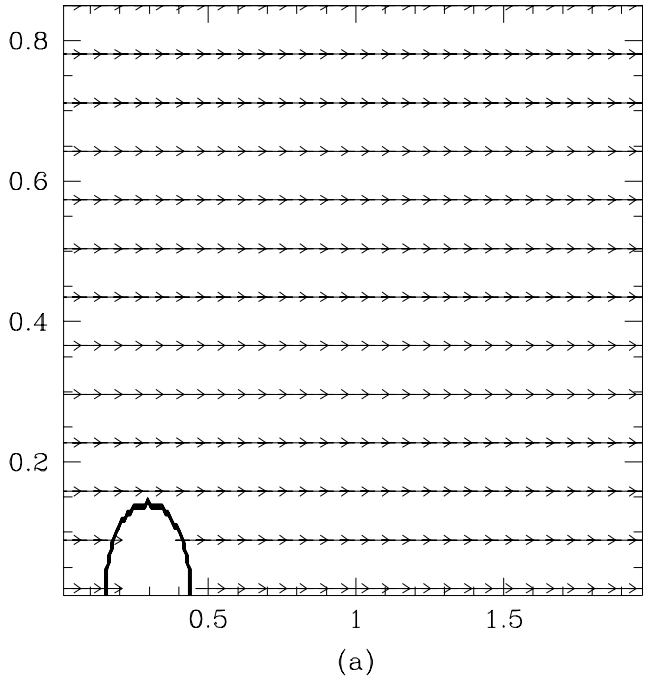
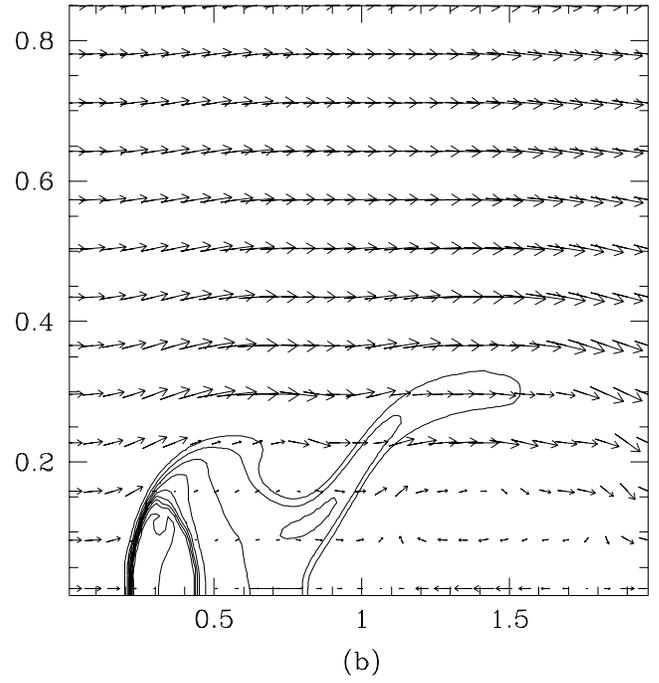
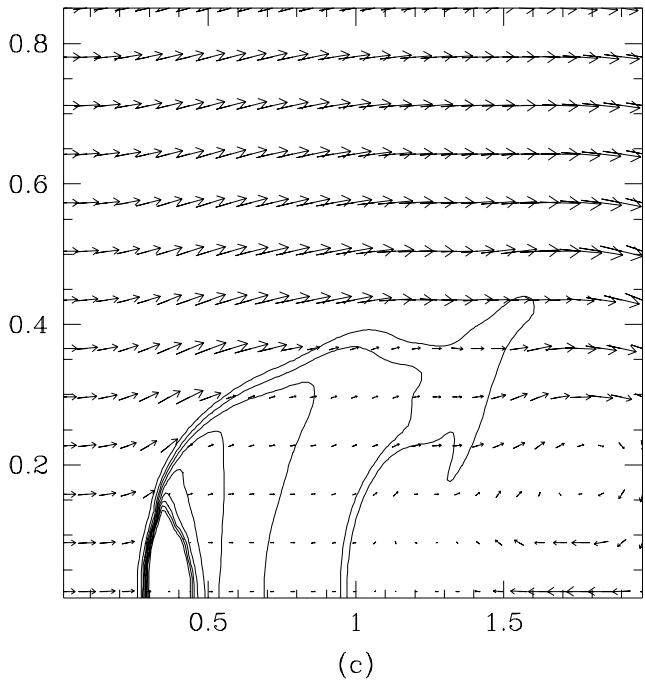
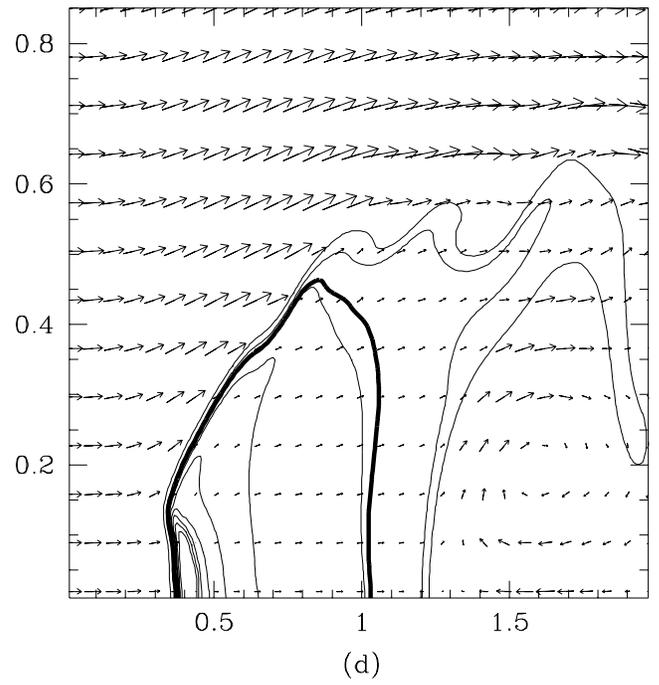
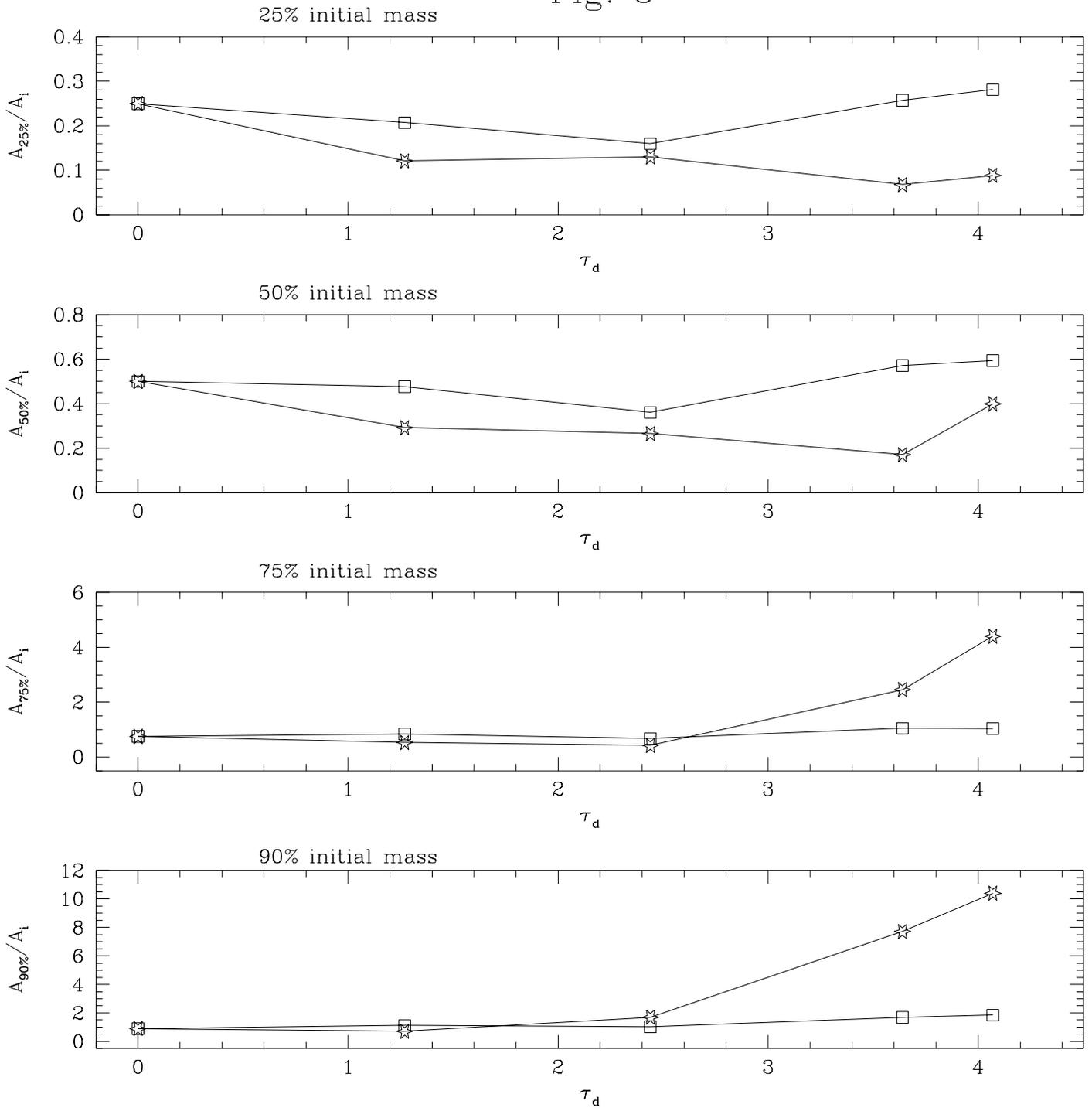
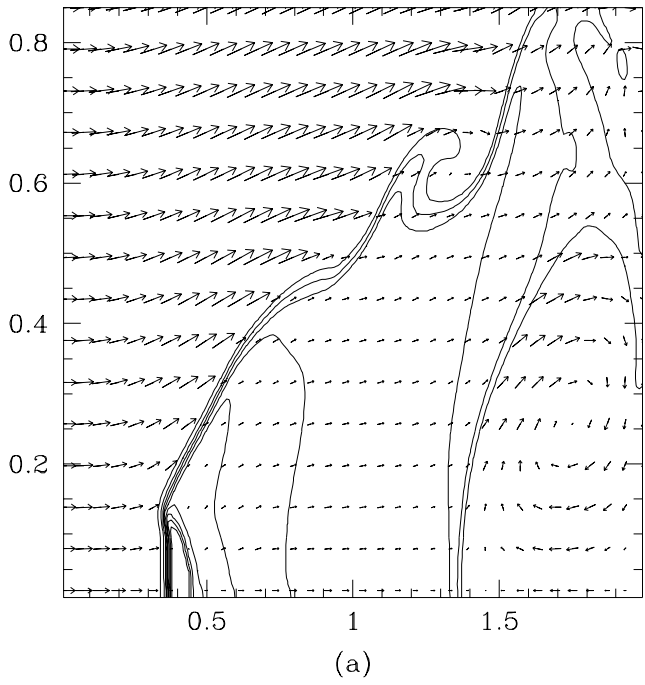
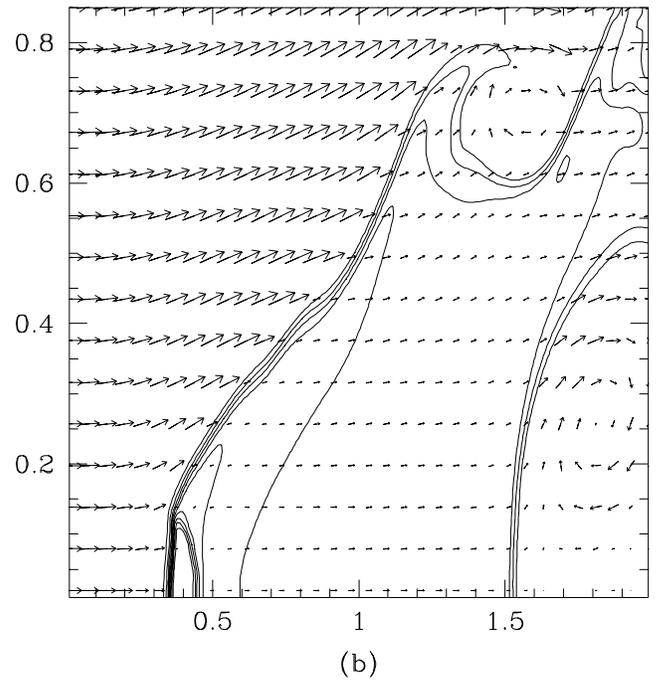
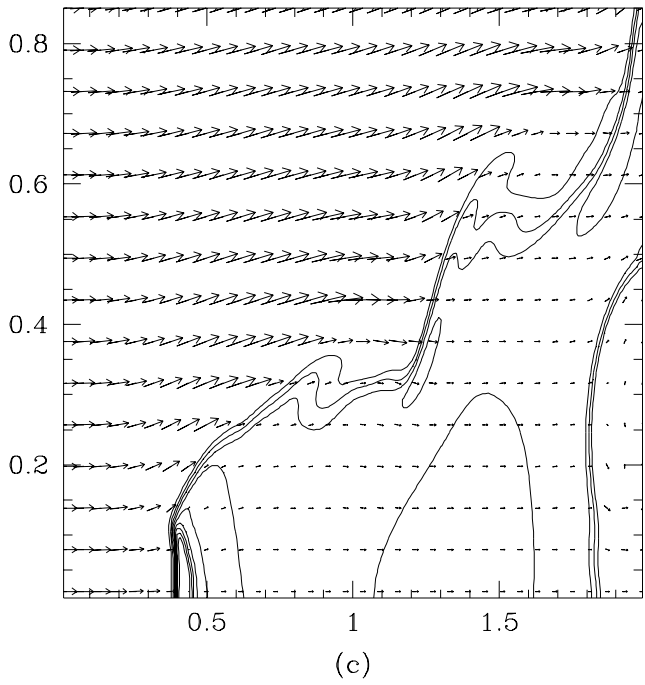
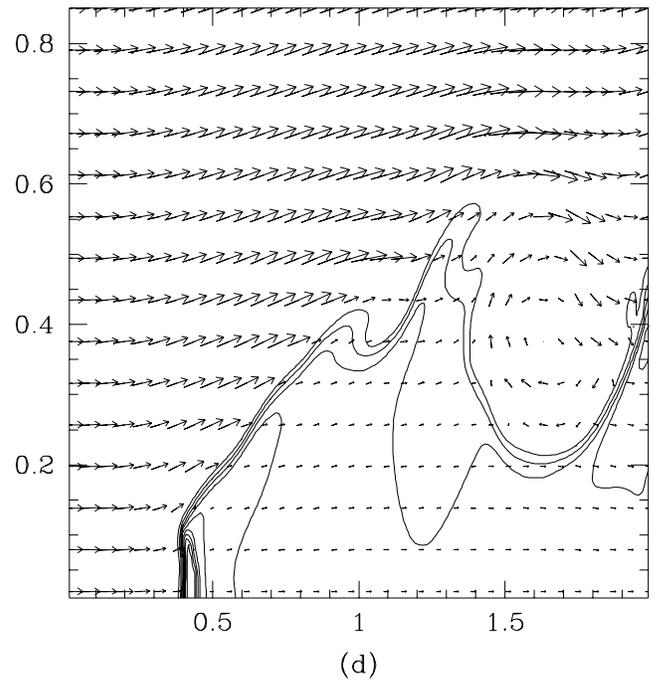
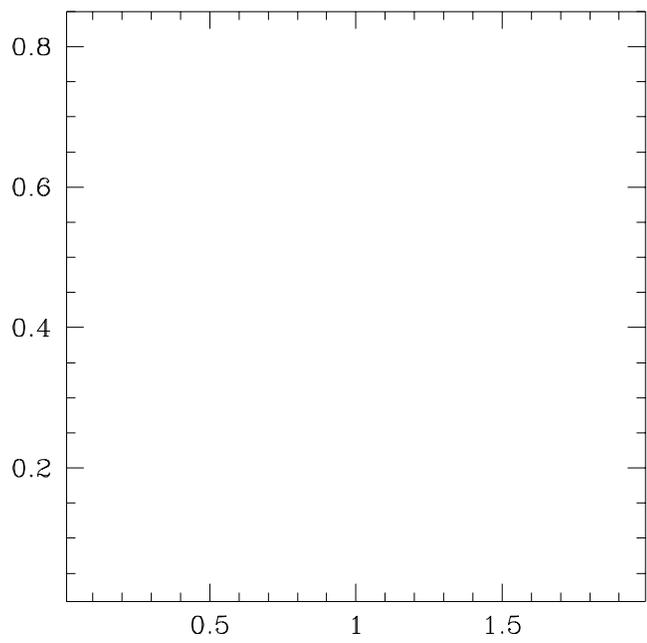
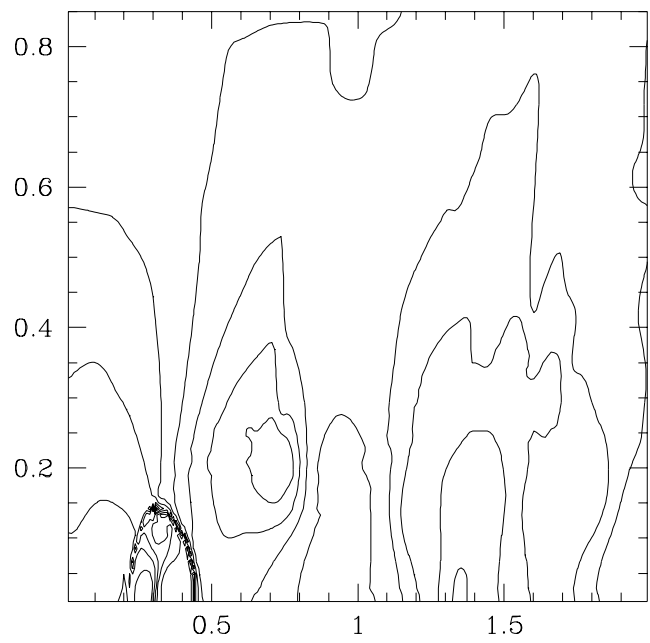
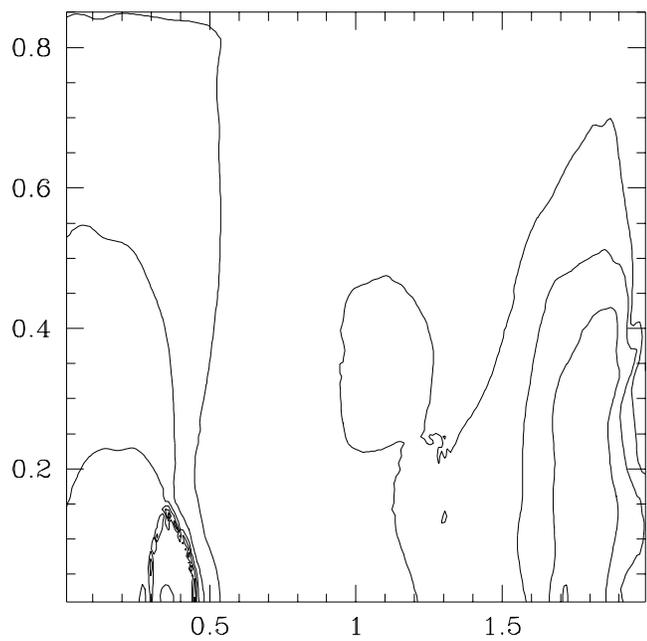
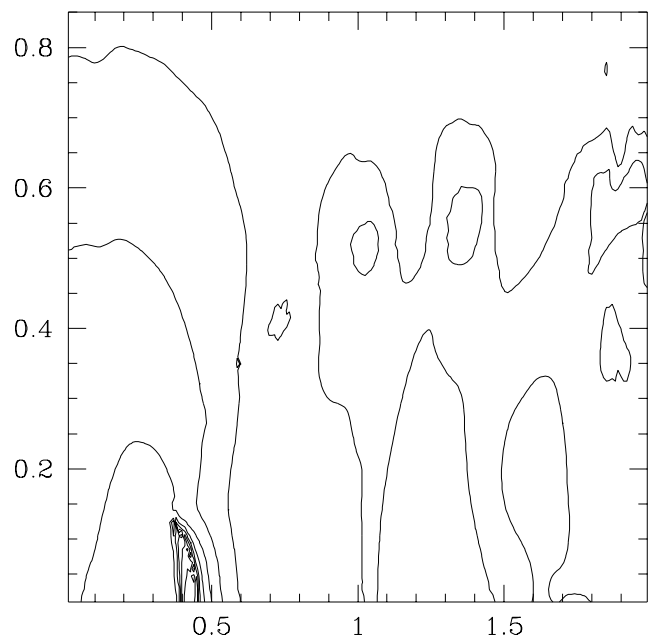
\end{document}